\documentclass[twocolumn,twoside,slac]{revtex4}
\usepackage{graphicx}
\usepackage{fancyhdr}
\begin{document}

\title{The Virtual Monte Carlo}

%

\author{I. H\v{r}ivn\'{a}\v{c}ov\'{a}}
\affiliation{IPN, Orsay, France}
\author{D. Adamov\'{a}}
\affiliation{NPI, ASCR, Rez, Czech Republic}
\author{V. Berejnoi, R. Brun, F.Carminati, A. Fass\`{o}, E. Fut\'{o}, A. Gheata, A. Morsch}
\affiliation{CERN, Geneva, Switzerland}
\author{I. Gonz\'{a}lez Caballero}
\affiliation{IFCA, Santander, Spain}
\author{\it{For the ALICE Collaboration}}

\begin{abstract}
The concept of Virtual Monte Carlo (VMC) has been developed by the ALICE 
Software Project to allow different Monte Carlo simulation programs to run 
without changing the user code, such as the geometry definition, the detector 
response simulation or input and output formats. Recently, the VMC classes 
have been integrated into the ROOT framework, and the other relevant packages 
have been separated from the AliRoot framework and can be used individually by 
any other HEP project. The general concept of the VMC and its set
of base classes provided in ROOT will be presented. Existing implementations for 
Geant3, Geant4 and FLUKA and simple examples of usage will be described.

\end{abstract}

\maketitle

\thispagestyle{fancy}


\section{Introduction}
The concept of Virtual Monte Carlo (VMC) has been gradually developed by the ALICE
Software project \cite{alice-off}. From the beginning, the ALICE collaboration
has adopted a strategy for the development of the simulation framework 
that would allow a smooth transition from the currently used transport code, 
Geant3 \cite{g3}, to new ones Geant4 \cite{g4} and Fluka \cite{fluka}.
Instead of maintaining the Geant3 based code written in FORTRAN and developing 
in parallel a new framework, based on a new simulation program,
the user code was gradually migrated from FORTRAN to C++ and a general
C++ interface to a transport Monte Carlo (MC) was developed.

The VMC development went through the following phases:

1. The C++ class, TGeant3, providing access to Geant3 data structures (common blocks)
and functions was introduced. This provided a starting point for a full migration
of the user code from FORTRAN to C++.

2. The abstract C++ class, AliMC, was defined as a generalization of TGeant3.
This gave the initial step for the development of the Geant4 interface and 
the explicit Geant3 dependencies in the user code were also taken away. 
However, the implementations of the AliMC interface for both Geant3 and Geant4 
were dependent on the ALICE software.

3. The interfaces to the user Monte Carlo application were
introduced. The dependence of the implementations of the AliMC interface 
on the ALICE software could then be removed and the VMC was also made available 
to non-ALICE users.

\section{Architecture}

\subsection{The VMC concept}
\begin{figure}[t]
\includegraphics[width=8cm]{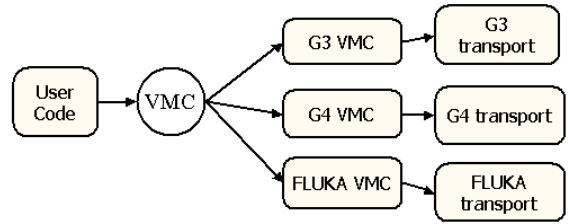}
\caption{The Virtual Monte Carlo concept.} \label{VMC-concept-now}
\end{figure}

With the VMC concept the user Monte Carlo application can be defined
independently of a specific transport code (see Fig.~\ref{VMC-concept-now}).
It can then be run with all supported Monte Carlos, without changing the user code, 
ie., the geometry definition, the detector response simulation and input 
or output formats. The selection of a concrete Monte Carlo (Geant3, Geant4 or Fluka)
is made dynamically at run time.

The VMC is based on the ROOT system \cite{root}, which is used mainly
for scripting and dynamical loading of libraries. Once the VMC application
has been defined, simulations can be run interactively from the Root UI or
using Root macros.

\subsection{Design}
\begin{figure}[t]
\includegraphics[width=8cm]{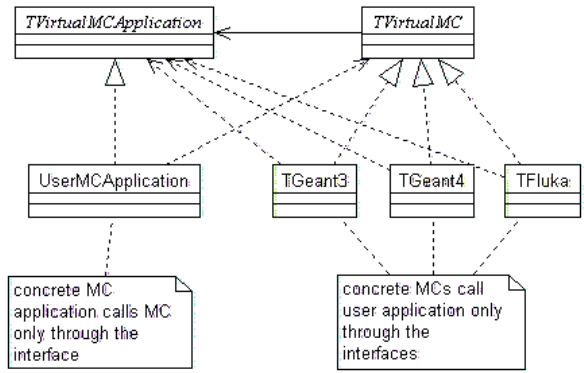}
\caption{The Virtual Monte Carlo design.} \label{design}
\end{figure}

In order to completely decouple the user code from the concrete
Monte Carlo, the interfaces to both the Monte Carlo itself and to the user 
application code have been introduced as shown in Fig.~\ref{design}.
In the following subsections, all interfaces will be discussed in detail.

\subsubsection{Virtual MC}
The Virtual MC interface (class TVirtualMC) was 
the first interface written and it is the most robust one.
It has been defined as a generalization of Geant3 functions for the definition
of simulation tasks and it provides:
\begin{itemize}
\item Methods for building and accessing geometry
\item Methods for building and accessing materials
\item Methods for setting physics
\item Methods for accessing transported particle properties during stepping
\item Methods for run control
\end{itemize}
The implementations of the Virtual MC for concrete transport programs
are part of the VMC distribution and are provided for the user.
At the present time, the Geant3 VMC and the Geant4 VMC are in distribution, 
the Fluka VMC will be available soon. 

\subsubsection{Virtual MC Application}
The Virtual MC Application interface (class TVirtualMCApplication) 
is the interface to a user application code. It defines user actions 
at each stage of a simulation run:
\begin{tabbing}
Armadillo: \=        \kill
\> Construct geometry\\
\> Init geometry\\
\> Generate primaries\\
\> Begin event\\
\> Begin primary\\
\> Pre Track\\
\> Stepping\\
\> Post Track\\
\> Finish primary\\
\> Finish event
\end{tabbing}
The implementation of the Virtual MC Application completely defines 
the user application and has to be provided by the user. 
 
\subsubsection{Virtual MC Stack}
The Virtual MC Stack interface (class TVirtualMCStack)
defines the interface to a user defined particles stack.
Users can choose one of the concrete stack classes provided
in the VMC examples or can implement their own stack class.

\subsubsection{Virtual MC Decayer}
The last interface in the VMC, the Virtual MC Decayer (class 
TVirtualMCDecayer), defines the interface to the external decayer.
The implementation of this interface by a user is optional.

\section{Use of VMC}
The user VMC application code is written by implementing the MC Application
class. In the case of very simple applications the user can write everything 
in the one class. In more complex cases it can be convenient to define the user application
class as a composition of more action classes, as is shown in the VMC examples
described in section 5.1. 

In this section, three examples of a user code based on the VMC will be given.

\subsection{Geometry construction}

In Example~\ref{table-geom-construction}, an example of a geometry definition is
given. The geometry is defined by calls to the Virtual MC interface. 
In the first block, a volume named ``TRTU'' of the shape tube of an inner 
radius of 0~cm and outer radius of 60~cm, is created and associated with a material
defined by the tracking medium identifier ``idAl''. In the second block,
this volume is placed at the position (-100~cm, 0~cm, 0~cm) in the mother volume
named ``EXPH''.

\begin{table*}[t]
\begin{center}
\caption{Example of a user code for geometry construction using the VMC}
\begin{tabular}{|l|}
\hline 
\\
void MyMCApplication::ConstructGeometry()                       \\
\{                                                              \\
\hskip 5mm \it{// Create tracker tube volume}                   \\
\hskip 5mm  Double\_t trackerTube[3];                           \\
\hskip 5mm  trackerTube[0] =  0.;                               \\ 
\hskip 5mm  trackerTube[1] = 60.;                               \\ 
\hskip 5mm  trackerTube[2] = 50.;                               \\ 
\hskip 5mm  gMC$->$Gsvolu("TRTU", "TUBE", idAl, trackerTube, 3);\\
\\
\hskip 5mm \it{// Place tracker tube volume}                    \\
\hskip 5mm  Double\_t posX = -100.;                             \\
\hskip 5mm  Double\_t posY =    0.;                             \\
\hskip 5mm  Double\_t posZ =    0.;                             \\
\hskip 5mm  gMC$->$Gspos("TRTU", 1, "EXPH", posX, posY, posZ, 0, "ONLY"); \\
\}                                                              \\
\\  
\hline
\end{tabular}
\label{table-geom-construction}
\end{center}
\end{table*}

The functions for geometry building are compatible with Geant3 (both by name 
and in the list of parameters). The VMC also uses the Geant3 system of default 
physical units. In future, using the Root geometrical modeller directly will 
be possible as an alternative to this Geant3 style. (For more details, 
see section 6.)

\subsection{Primary particles}
In Example~\ref{table-primary-particles}, an example of how to define primary 
particles is given. This is done by calls to the Virtual MC Stack interface. 
The particle type (proton, electron, ...) is defined using
the PDG encoding and particle static properties (mass, charge, ...)
are taken from the particle database in ROOT (represented by the TDatabasePDG 
class).

\begin{table*}[t]
\begin{center}
\caption{Example of a user code for primary particles definition using the VMC}
\begin{tabular}{|l|}
\hline 
\\
void MyMCApplication::GeneratePrimaries()              \\
\{                                                     \\
\hskip 5mm \it{// Define particle properties: }        \\
\hskip 5mm \it{// PDG encoding: pdg           }        \\ 
\hskip 5mm \it{// position: vx, vy, vz, t     }        \\ 
\hskip 5mm \it{// momentum: px, py, pz, e     }        \\ 
\hskip 5mm \it{// ...                         }        \\
\\
\hskip 5mm \it{   // Add particle to MC stack }        \\
\hskip 5mm gMC$->$GetStack()                           \\
\hskip 18mm      $->$SetTrack(toBeDone, -1, pdg, px, py, pz, e, vx, vy, vz,t, ...);\\
 \}                                                    \\
\\  
\hline
\end{tabular}
\label{table-primary-particles}
\end{center}
\end{table*}

\subsection{Detector response}
In Example~\ref{table-detector-response}, an example of a user stepping function
is shown. This function is called by MC at each step. In this example
the properties of the particle transported are obtained via calls to
the Virtual MC interface and then saved in the user own hits objects.
For large detectors it is recommended to delegate this function to 
stepping functions defined in subdetector classes.

\begin{table}[t]
\begin{center}
\caption{Example of a user code for a detector response simulation using the VMC}
\begin{tabular}{|l|}
\hline 
\\
void MyMCApplication::Stepping()                       \\
\{                                                     \\
\hskip 5mm \it{// Get current volume ID}               \\
\hskip 5mm  Int\_t copyNo;                             \\
\hskip 5mm  Int\_t id = gMC$->$CurrentVolID(copyNo);   \\
\\
\hskip 5mm \it{// Check if step is performed in the sensitive volume } \\
\hskip 5mm if (id != fSensitiveVolumeID) return false; \\
\\
\hskip 5mm \it{// Get track position}                  \\
\hskip 5mm  Double\_t x, y, z;                         \\ 
\hskip 5mm  gMC$->$TrackPosition(x, y, z);             \\ 
\\
\hskip 5mm \it{// Get energy deposit}                  \\ 
\hskip 5mm  Double\_t edep = gMC$->$Edep();            \\
\\
\hskip 5mm \it{   // Create user hit}                  \\
\hskip 5mm  mySD$->$AddHit(x, y, z, edep);             \\
 \}                                                    \\
\\  
\hline
\end{tabular}
\label{table-detector-response}
\end{center}
\end{table}

\section{Available MCs}

In this section we will outline the present status and give a short description 
of the implementations of the Virtual MC for the three transport MCs: 
Geant3, Geant4 and Fluka.

\subsection{Geant3 VMC}
The Geant3 program \cite{g3} was written to describe the passage of elementary 
particles through matter. Originally designed for high energy physics 
experiments, it has also found applications outside this domain in the areas 
of medical and biological sciencies, radioprotection and astronautics. 
The first version was released in 1974 and the system was 
developed with some continuity over 20 years till the last release 3.21 in 1994. 
It has become a popular and widely used tool in the HEP community.

The Geant3 VMC, which implements the Virtual MC interface to the Geant3 program,
is provided within a single package ``geant3'' together with Geant321 itself.
This ``geant3'' package is available from the ROOT Web site \cite{root}.

As the Virtual MC was largely inspired by Geant3, its implementation
for Geant3 was straightforward and has no limitations.

Besides the implementation of the Virtual MC, the Geant3 VMC also includes
the Geant3 Geometry Browser \cite{g3gui}, a GUI which provides a variety
of functions, namely:
\begin{itemize}
\item visualization of the geometry volumes tree
\item drawing of volumes and interactive setting of drawing options
\item browsing material and tracking medias parameters
\item browsing applied cuts and activated physics processes
\item plotting of dE/dx and cross-sections for a selected physical process
\end{itemize}
The implementation is based on the Root GUI classes.  

\subsection{Geant4 VMC}

The Geant4 project \cite{g4} was started in 1994, the first production version was 
released in 1998 and the system is continuously under developement by the Geant4 
Collaboration. Its areas of application include particle and nuclear physics 
experiments, medical, accelerator and space physics studies.

The Geant4 VMC, which implements the Virtual MC interface for the Geant4 program,
is provided within the ``geant4\_vmc'' package and requires a prior Geant4 
installation. This ``geant4\_vmc'' package is available from the ROOT Web 
site \cite{root}.

The implementation of the interface to MC for Geant4 was presented 
at the CHEP 2001 conference \cite{chep2001}. The design, the implementation 
and also the problems arising from the G3toG4 approach and their foreseen 
solutions were discussed. Despite improvements and design changes, 
the structure and the components of the package presented there 
can be found in the current Geant4 VMC package. The major change applied
since then was that the dependencies on the ALICE classes in AliGeant4 
have been replaced by the dependencies on the interfaces to a user 
application. This meant that all classes from AliGeant4 could be moved to 
the experiment independent part, TGeant4, that has been then renamed
Geant4 VMC.   

The main improvement from that time was in minimizing the limitations 
of the G3toG4 tool in the Geant4 4.0 release. New classes to support 
the reflection symmetry and also a limited support for the ``MANY'' 
volumes positions have been introduced. The G3toG4 tool does not resolve positions with ``MANY'' 
automatically, however a user can specify the overlapping volumes using the 
G4gsbool() function and then the overlaps for these volumes and their daughters
are automatically resolved using Boolean solids. The volume 
with a ``MANY'' position can only have this position. The corresponding
function Gsbool() has been added to the Virtual MC interface.

The VMC interface provides a common denominator for all implemented MCs
and cannot cover all commands available in a Geant4 user session
through the Geant4 UI. Switching between the Root UI and the Geant4 UI 
gives the VMC user the possibility of working with the native Geant4 UI 
when needed or desired. It is also possible to process a foreign command or a 
foreign macro in both UIs, for example the Root commands and macros 
can be processed in the Geant4 UI and vice versa. 

In a similar way as the Geant3 VMC, the Geant4 VMC also includes the Geant4 
Geometry Browser. It was implemented in an analogous way. It provides 
the same functionality for browsing geometry. It does not include the panels
that allow to browse the activated physical processes
and their characteristics.

In addition, the Geant4 VMC also includes an XML convertor. With this convertor
the Geant4 geometry can be exported to XML in the AGDD format \cite{agdd} and then 
browsed and visualized using the GraXML tool \cite{graxml}. The XML 
convertor classes are independent of the Geant4 VMC, they can be compiled 
in a separate library and also used with a native Geant4 application.

\subsection{Fluka VMC}
The history of FLUKA \cite{fluka} goes back to 1962-1967. Over the years it went 
through three different generations, which can be roughly identified
as the FLUKA of the '70s, the FLUKA of the '80s and today's FLUKA as a fully 
integrated particle physics Monte Carlo simulation package. It has many applications 
in high energy experimental physics, engineering, shielding, detector and telescope 
design, cosmic ray studies, dosimetry, medical physics and radio-biology. 
       
The Fluka VMC is now under development within the ALICE collaboration and 
the FLUKA team. The development version of the VMC implementation is 
in the TFluka package of AliRoot. 

Most of the functionality required by the VMC is
already fully operational:
\begin{itemize}
\item Functions for building and accessing geometry (through Flugg) 
\item Functions for accessing transported particle properties during stepping
\item Recording particles in the VMC stack
\item Functions for run management
\end {itemize}
Under development are:
\begin{itemize}
\item Functions for selecting activated physics processes
\item Interface to the external decayer
\end {itemize}

The geometry part of the Fluka VMC is implemented with the use of
the Geant4 VMC and the Flugg tool \cite{flugg}. Flugg  was developed 
a few years ago so that the Geant4 geometry could be used directly in the 
Fluka particle transport. In the near future this part will be replaced 
by the Root geometrical modeller. (For more details, see section 6.)

\section{Examples}

\subsection{Examples provided with the VMC}
To demonstrate the use of the VMC, Geant4 novice examples N01, N02 and 
N03 were rewritten in the VMC framework.
The Geant4 novice examples were chosen in order to show
the similarities and differences between the two frameworks.
Having in parallel, both the Geant4 VMC and the Geant4 native application, 
allows to verify the consistence of the results and to compare
performancies. 

The VMC examples demonstrate the implementation of the user 
MC application and MC stack classes. While in the first example all 
functions of the MC application are implemented directly in the Ex01MCApplication
class, in the other examples the MC application class is defined as 
being a composition of more action classes (e.g  the detector construction class)
and it delegates most of its functions to its components.

\begin{table*}[t]
\begin{center}
\caption{Root macros, run\_g3.C and run\_g4.C, that show how to run the VMC example E01. 
MC specific parts are given in the parallel columns.}
\begin{tabular}{|l|l|}
\hline 
\multicolumn{2}{|l|}{ \hskip 20mm \{                              }   \\
\multicolumn{2}{|l|}{ \hskip 25mm \it{// Load basic libraries}    }   \\
\multicolumn{2}{|l|}{ \hskip 25mm gROOT$->$LoadMacro("basiclibs.C");} \\ 
\multicolumn{2}{|l|}{ \hskip 25mm basiclibs();                    }   \\ 
\multicolumn{2}{|l|}{ \hskip 25mm                                 }   \\

\hskip 5mm // Load Geant3 libraries       & 
\hskip 5mm // Load Geant4 libraries                                   \\ 
\hskip 5mm gROOT$->$LoadMacro("g3libs.C"); & 
\hskip 5mm gROOT$->$LoadMacro("g4libs.C");                            \\                       
\hskip 5mm g3libs();                      & 
\hskip 5mm g4libs();                                                  \\
\multicolumn{2}{|l|}{ \hskip 25mm                                }    \\
\multicolumn{2}{|l|}{ \hskip 25mm \it{// Load this example library}}  \\                     
\multicolumn{2}{|l|}{ \hskip 25mm gSystem$->$Load("libexample01"); }  \\
\multicolumn{2}{|l|}{ \hskip 25mm                                }    \\

\multicolumn{2}{|l|}{ \hskip 25mm  \it{// MC application}        }    \\
\multicolumn{2}{|l|}{ \hskip 25mm  Ex01MCApplication* appl       }    \\            
\multicolumn{2}{|l|}{ \hskip 30mm = new Ex01MCApplication("Example01", "The example01 MC application"); } \\

\multicolumn{2}{|l|}{ \hskip 25mm                                }   \\
\hskip 5mm  appl$->$InitMC("g3Config.C"); &  
\hskip 5mm  appl$->$InitMC("g4Config.C");                            \\

\multicolumn{2}{|l|}{ \hskip 25mm                                }   \\
\multicolumn{2}{|l|}{ \hskip 25mm appl$->$RunMC(1);              }   \\
\multicolumn{2}{|l|}{ \hskip 20mm \}                             }   \\
\multicolumn{2}{|l|}{ \hskip 25mm                                }   \\
\hline 
\end{tabular}
\label{table-run-macros}
\end{center}
\end{table*}

All examples are executed by processing the provided Root macros.
The macros for running the example E01 with Geant3 and Geant4 (run\_g3.C and run\_g4.C)
are shown in Example~\ref{table-run-macros}. In both macros, all necessary
libraries are first dynamically loaded, then the user MC application is created 
and initialized with a MC specific configuration macro (g3Config.C or g4Config.C) 
shown in Example~\ref{table-config-macro}. After initialization, a simulation run with 
the chosen concrete MC is executed for the specified number of events.

\begin{table*}[t]
\begin{center}
\caption{Root configuration macros: g3Config.C and g4Config.C}
\begin{tabular}{|l|}
\hline 
\textbf{g3Config.C:} \\                  
\\
void Config()                                          \\
\{                                                     \\
\hskip 5mm \it{// Geant3 VMC}                          \\
\hskip 5mm new TGeant3("C++ Interface to Geant3");     \\
\}                                                     \\
\\  
\hline
\textbf{g4Config.C:} \\
\\                  
void Config()                                               \\
\{                                                          \\
\hskip 5mm \it{// Run Configuration for Geant4}             \\
\hskip 5mm TG4RunConfiguration* runConfiguration = new TG4RunConfiguration();  \\
\\
\hskip 5mm \it{// Geant4 VMC}                               \\
\hskip 5mm new TGeant4("TGeant4", "The Geant4 Monte Carlo", runConfiguration);\\
\}                                                          \\
\\  
\hline
\end{tabular}
\label{table-config-macro}
\end{center}
\end{table*}

\subsection{AliRoot}
AliRoot is the ALICE off-line framework for simulation,
reconstruction and analysis. The simulation in AliRoot is fully based 
on the VMC. Hence the AliRoot framework can be given as a complex example 
of a VMC application. 

The AliRoot framework will not be discussed in more detail in this paper,
but the reader is directed to the other CHEP03 conference papers
related to this subject: ``The AliRoot Framework, status and perspectives''
\cite{aliroot-chep03}, ``Simulation in ALICE'' \cite{simulation-chep03}, 
``ALICE experience with Geant4'' \cite{g4alice-chep03}.

\begin{figure}[t]
\includegraphics[width=8cm]{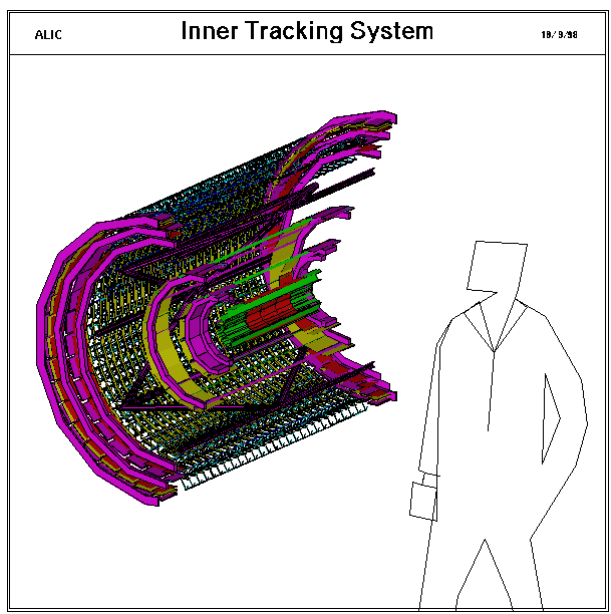}
\caption{Geometry for ITS detector drawn with Geant3.} 
\label{aliroot-its-geometry}
\end{figure}

\begin{figure}[t]
\includegraphics[width=8cm]{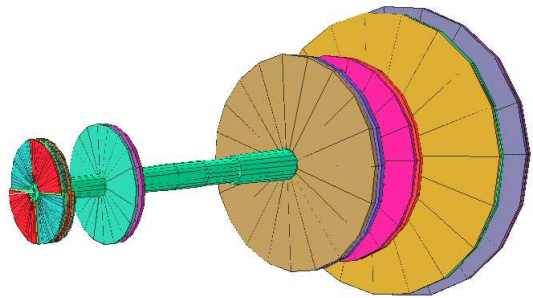}
\caption{Geometry for the Dimuon Arm spectrometer drawn with Geant4 and GraXML.} 
\label{aliroot-muon-geometry}
\end{figure}

\begin{figure}[t]
\includegraphics[width=8cm]{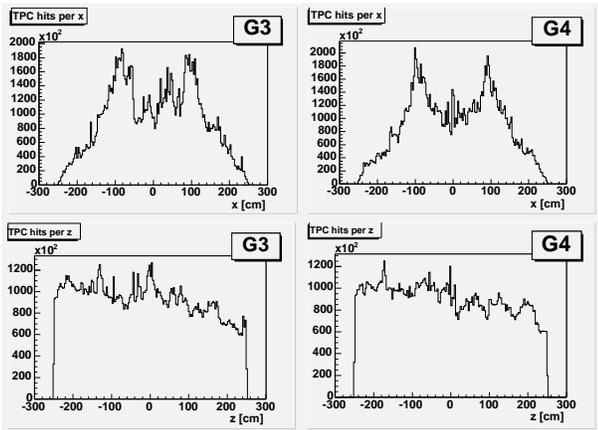}
\caption{The x,z-distributions of hits in the TPC detector.} 
\label{aliroot-hits}
\end{figure}

In the following, a few figures demonstrating the use of AliRoot with two MCs, 
Geant3 and Geant4, will be given. In Fig.~\ref{aliroot-its-geometry} and  
Fig.~\ref{aliroot-muon-geometry} the geometries for two ALICE detector 
subsystems (the ITS and the Dimuon arm) are shown - the first drawn with Geant3, 
the second drawn with Geant4 and GraXML. In Fig.~\ref{aliroot-hits}, 
the x and z distributions of hits in the TPC subsystem are shown for two 
transport MCs: Geant3 and Geant4, giving qualitatively similar results.  

\section{Future}
Over the past two years the ALICE Offline project, in close collaboration with
the ROOT team, has developed a new multi-purpose geometrical modeller for HEP.
It is called TGeo and has been included in ROOT \cite{tgeo-chep03}. 
This new geometrical modeller is currently being integrated into the VMC.

The VMC user will then have also the possibility
to define geometry directly using the new geometrical modeller. The current interface
will be kept for backward compatibility. Apart from making a geometry construction 
via the VMC more user-friendly, this integration will also improve 
the performance of the MC simulation. The new modeller has been optimised 
for performance and runs faster than Geant3 for most geometries tested. 

For Geant3 and Fluka, the integration of TGeo is being made by replacing
the native geometry with the TGeo one. The new version of
Geant3 with the new Root geometry package is close to completion and
the work on the integration of TGeo into Fluka is already well 
advanced.

For Geant4, the object-oriented methodology can be exploited and TGeo can 
be integrated with Geant4 through an abstract interface to the Geant4 geometry.
At present, the Geant4 geometry navigator class is not based on an abstract
interface, its generalization will be required to allow an alternative
implementation using TGeo. Discussions with the Geant4 team have started 
and the design and prototype implementation for the abstract navigator
and transportation are now part of the Geant4 developments planned for the 6.0
release \cite{g4plans}.

In the meantime, a convertor from Root geometry to Geant4 native
geometry, RootToG4, has been developed. It is already operational and close
to completion. It will allow the migration from the current VMC geometry definitions
to TGeo, before the solution via the abstract navigator is available.

\begin{figure}[t]
\includegraphics[width=8cm]{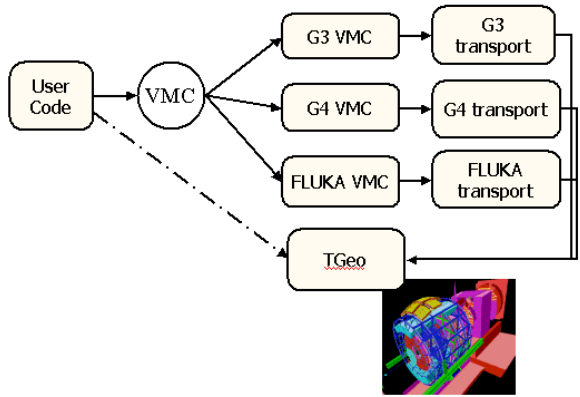}
\caption{The Virtual Monte Carlo and TGeo integration.} 
\label{VMC-concept-future}
\end{figure}

The new concept of the VMC including the Root geometrical modeller
is shown in Fig.~\ref{VMC-concept-future}. 

\section{Distribution}
The VMC is distributed with the ROOT system \cite{root}.
It consists of the following packages:
\begin{itemize}
\item mc: the core package (interfaces)
\item geant3: Geant321 + Geant3 VMC
\item geant4\_vmc: Geant4 VMC
\item examples
\end{itemize}
The ``mc'' package is directly included in ROOT, ``geant3'' and ``geant4\_vmc''
are available from the ROOT CVS server \cite{rootcvs} as independent modules 
and the package with the examples is provided within ``geant4\_vmc''.

The tarballs with sources are also available from the VMC
Web page \cite{vmc}. 

\section{Conclusions}
The Virtual Monte Carlo provides a simulation framework which is independent 
of any concrete MC and is based on the ROOT system. Geant3 and Geant4 have 
already been integrated and the same work with Fluka is in progress.

The main advantage with VMC is that the user 
can run the same simulation program with three different transport MCs.
The obvious consequence is that the different models can be compared and
better understood.
VMC also facilitates the use of less user-friendly tools such as Geant3 and Fluka
and - being inspired from Geant3 - it is also suitable for users starting 
with existing Geant3 applications.
  
The integration of the Root geometrical modeller will give the user
a new means for geometry definition, browsing, 
visualization and also verification directly within the scope of the VMC.

\begin{acknowledgments}
The authors wish to thank the Geant4 and Fluka teams for the constructive
collaboration and their help provided during this work.
\end{acknowledgments}


\end{document}